\documentclass{birkjour}
\usepackage{amssymb,eucal,bm,cite,mathtools}
\usepackage{tikz}
\usetikzlibrary{calc}

\title[Differential versus improved Epstein--Glaser renormalization]
{Systematic Renormalization at all Orders\\
in the DiffRen and Improved Epstein--Glaser\\
Schemes}

\author{Jos\'e M. Gracia-Bond\'ia}

\address{%
Department of Theoretical Physics\\
Universidad de Zaragoza\\
50009 Zaragoza, Spain;\\[2ex]
BIFI Research Center\\
Universidad de Zaragoza\\
50018 Zaragoza, Spain;\\[2ex]
Institute for Theoretical Physics\\
Universidad Aut\'onoma\\
28049 Madrid, Spain.}

\date{\today}

\newcommand{\bt}{\beta}            
\newcommand{\Dl}{\Delta}           
\newcommand{\dl}{\delta}           
\newcommand{\eps}{\varepsilon}     
\newcommand{\Ga}{\Gamma}           
\newcommand{\ga}{\gamma}           
\newcommand{\vf}{\varphi}          

\newcommand{\sP}{\mathcal{P}}      

\newcommand{\bR}{\mathbb{R}}       

\newcommand{\xx}{\bm{x}}           

\DeclareMathOperator{\grad}{grad}  

\newcommand{\bull}{\mathord{\scriptstyle\bullet}} 
\newcommand{\del}{\partial}        
\newcommand{\dsp}{\displaystyle}   
\newcommand{\downto}{\downarrow}   
\newcommand{\Rbar}{{\mkern2mu \overline{\mkern-2mu R}}} 
\newcommand{\w}{\wedge}            
\newcommand{\x}{\times}            
\def\duo<#1,#2>{\langle#1,#2\rangle} 

\newcommand{\set}[1]{\{\,#1\,\}}   
\newcommand{\word}[1]{\quad\mbox{#1}\quad} 

\newcommand{\copadevino}{{\!{\raisebox{-2mm}{%
\begin{tikzpicture}[scale=0.35]
\coordinate (A) at (0,0) ; \coordinate (B) at (1,0) ; 
\coordinate (C) at (0.5,-0.8) ;
\draw ($ (A)!-0.3!(C) $) -- ($ (A)!1.3!(C) $) ;
\draw ($ (B)!-0.3!(C) $) -- ($ (B)!1.3!(C) $) ;
\draw (A) parabola[bend pos=0.5] bend +(0,0.2) (B) ;
\draw (A) parabola[bend pos=0.5] bend +(0,-0.2) (B) ;
\foreach \pt in {A,B,C} \draw (\pt) node{$.$} ;
\end{tikzpicture}}}\!\!}} 

\begin{document}

\begin{abstract}
Proceeding by way of examples, we update the combinatorics of the
treatment of Feynman diagrams with subdivergences in differential
renormalization from more recent viewpoints in Epstein--Glaser
renormalization in $x$-space.
\end{abstract}

\maketitle

\section{Introduction}
\label{sec:ad-altare}

Ultraviolet divergent amplitudes of Feynman diagrams in $x$-space are
given by functional expressions too singular at short distances to
define distributions on all of space(-time). Quite often, however,
they can be represented on their domain of definition as derivatives
of \textit{bona fide} distributions, and so extended to the whole
space. ``Differential renormalization'' (DiffRen for short) in this
sense does not need previous regularization steps. It was introduced
by Freedman, Johnson and Latorre in a path-breaking
article~\cite{FreedmanJL92}.

While that reference is a true \textit{tour de force}, showing the
calculational power of the method (and of its authors), it did not
attempt to disentangle the combinatorics of successive
renormalizations for non-primitive graphs in a systematic way. This
task was taken up in later work in differential
renormalization~\cite{LCristinaMVV94}.

Differential renormalization remained popular during the nineties, to
peter out in the third millennium. Now, a spate of papers
\cite{NikolovST14,Elara,DuetschFKR14} have recently dealt with
renormalization in $x$-space in the spirit of the classic article by
Epstein and Glaser~\cite{EpsteinG73}, whose kinship with DiffRen is
evident. Each of those solves the renormalization recursion in its own
way. Among them, reference~\cite{Elara} adheres to DiffRen closely and
improves it. Our aim here is to pedagogically revisit from its
viewpoint the combinatorics of the ``subtraction'' of subdivergences
in~\cite{LCristinaMVV94} as well, comparing methods and (hopefully)
bringing improvement again.

As in the last-mentioned paper, for simplicity we deal mostly with the
Euclidean massless $\phi^4_4$~model.

\section{The importance of degree}
\label{sec:al-huerto}

We start by a point of rigour, that some readers may wish to skip.
Naturally the first step in \cite{LCristinaMVV94} is to
determine whether a given graph needs renormalization. The answer they
give, in view of Weinberg's ``power counting''
theorem~\cite{youngfirebrand,Manoukian}, is to find the superficial
degree of each generalized vertex (subgraph) in the graph. Now, that
theorem yields a \textit{sufficient} condition for renormalizability.
While the question of improving on it is moot for scalar models,
experience with renormalization of massless graphs suggests a less
restrictive criterion. To wit, extension of log-homogeneous graphs in
the sense of \cite{NikolovST14,Elara} produces log-homogeneous graphs
of the same order and higher degree. Whenever the extension can be
made to a log-homogeneous graph of the \textit{same} degree, we
understand that we deal with a matter of (re)definition, the diagram
is convergent, and no renormalization has taken place.

The subject is discussed in~\cite{NikolovST14}, but we deem it
worthwhile to bring it here for completeness in our review
of~\cite{LCristinaMVV94}. We show a truly trivial example in the
Minkowski space~$M_4$. The distribution $\dl(x^2)$ appears routinely
in formulas for the propagators of free massless fields. However, its
meaning is not altogether obvious. Given any distribution $f$
on~$\bR$, one is able to define pullback distributions $f(P)$ on
smooth hypersurfaces $P$ (codimension 1 submanifolds) of~$M_4$; in
particular one defines $\dl(P)$. However the lightcone $x^2 = 0$ is
not smooth at the origin, where $\grad x^2$ vanishes. Thus
\textit{prima facie} $\dl(x^2)$ is defined on $M_4 \setminus \{0\}$
only ---as a homogeneous distribution of order~$-2$. Consider,
however, for small positive~$\eps$, the distributions:
$$
\duo< \dl_\pm(x^2 - \eps), \phi> = \int_\pm \phi \,\mu_\eps,
$$
where the integrals are respectively concentrated on the upper and
lower sheet of the hyperboloid $\set{x : x^2 = \eps}$, and $\mu_\eps$
is a Leray form such that $dx^2 \w \mu_\eps = d^4x$. One can take
$\dsp \mu_\eps = \frac{d^3\xx}{2t(r;\eps)} :=
\frac{d^3\xx}{2\sqrt{r^2 + \eps}}$ with $r \equiv |\xx|$. Hence,
$$
\duo< \dl_\pm(x^2 - \eps), \phi> = \frac{1}{2} \int
\frac{\phi(\pm\sqrt{r^2 + \eps},\xx)}{\sqrt{r^2 + \eps}} \,d^3\xx
= 2\pi \int_0^\infty
\frac{\bar\phi(\pm\sqrt{r^2 + \eps},r)}{\sqrt{r^2 + \eps}}\, r^2\,dr.
$$
We have called $\bar\phi(t,r)$ the average value of $\phi$ on a sphere
of radius $r$ in $\xx$-space. Then the limits as $\eps \downto 0$,
\begin{equation}
\duo< \dl_\pm(x^2), \phi> 
= 2\pi \int_0^\infty \bar\phi(\pm r,r)\, r\,dr,
\label{eq:delta-cone} 
\end{equation}
are obviously well defined, and they extend the previous~$\dl(x^2)$.

The extension of $\dl(x^2)$ of course is not unique; but crucially the
one just defined preserves Lorentz symmetry and keeps the \textit{same}
degree of homogeneity: we then reckon that no ``renormalization'' has
taken place.%
\footnote{The ``original sin'' still shows in that, whereas for
$\eps > 0$ we can indefinitely apply to $\dl_\pm(x^2 - \eps)$ the
usual derivation rules, the expression $\dl'(x^2)$ on $M_4$ is
meaningless.}

A more sophisticated example is found in~\cite[Sect.~5]{NikolovST14}.
The propagator of the free electromagnetic field is the quotient of a
harmonic polynomial of degree~$2$ by the third power of $x^2$, and
thus is logarithmically divergent by power counting. However, it does
possess an homogeneous extension, therefore it is convergent in our
more precise sense. The example offers little doubt, since, as is well
known, it boils down to second derivatives of the massless
propagator~\cite{Stora05}.

\section{Dealing with a three-point problem}
\label{sec:al-kokumi}

Let us continue by pointing to another difference between DiffRen as
practiced in~\cite{FreedmanJL92,LCristinaMVV94} and the
(improved) Epstein--Glaser method~\cite{Carme}. In the former papers one
finds as extension for the ``fish'' graph of the $\phi^4_4$ model,
with vertices $(0,x)$:
$$
r_x[x^{-4}] = -\frac{1}{2}\,
\Dl\biggl( x^{-2} \log\frac{|x|}{l} \biggr),
$$
in the understanding that $l = 1/M$, where $M$ is their mass scale;
while, as explained in~\cite{Elara}, we prefer
$$
R_x[x^{-4}] = -\frac{1}{2}\, \biggl[
\Dl\biggl( x^{-2} \log\frac{|x|}{l} \biggr) - \pi^2\,\dl(x) \biggr],
$$
because $x^2 R_x[x^{-4}] = x^{-2}$, while $x^2 r_x[x^{-4}]$ fails to
reproduce the convergent amplitude~$x^{-2}$. The algebra property
of~\cite{Carme} generalizes throughout into the ``causal factorization
property'' of \cite{NikolovST14}, which is the basis for a streamlined
proof of the recursive renormalization of subdivergences.

Now, our lodestone to deal with recursive renormalization
in~\cite{Elara} was a rule contained in the very illuminating
paper~\cite{KuznetsovTV96}. It is written:
\begin{equation}
\duo< R[\Ga], \vf> = \duo< R[\ga], (\Ga/\ga)\,\vf>.
\label{eq:in-annum} 
\end{equation}
In this formula any subgraph $\ga$ of a given graph $\Ga$ is
identified as a subset of the set of vertices of $\Ga$ and the set of
\emph{all} lines joining any two elements of~this subset. There
$R[\Ga]$, $R[\ga]$ and~$\Ga/\ga$ denote amplitudes, respectively for
the renormalized graph and subgraph, and the bare cograph~$\Ga/\ga$;
the test function $\vf$ is supported outside the singular points of
the latter. The rule works as a necessary and sufficient
\textit{prescription}: the $R[\ga]$ are supposed known, and then
$R[\Ga]$ must conform to the formula above. It subsumes (the Euclidean
version of) the causal factorization property, which allowed a
streamlined proof of recursive renormalization in~\cite{NikolovST14},
but would be awkward in actual computation. Later we will show how
rule \eqref{eq:in-annum} works when there are internal vertices
in~$\Ga$, by means of an interesting six-loop diagram considered
in~\cite{LCristinaMVV94}.

In \cite{LCristinaMVV94}, on the other hand, it is claimed that
recursive renormalization is effected by use of Bogoliubov's
subtraction operators~\cite[Ch.~29]{BogoItself}; another good
reference for this method is~\cite{Manoukian}. For instance, according
to them, the bare graph $x^{-4}$, with vertices $x,0$, is renormalized
by
$$
r_{x,0}[x^{-4}] = (I - T_{x,0}) x^{-4}
\word{with} T_{x,0}\,x^{-4} := x^{-4} - \biggl(
- \frac{1}{2}\,\Dl\Bigl( x^{-2} \log \frac{|x|}{l} \Bigr) \biggr).
$$
Actually $T_{x,0}\,x^{-4}$ as written above makes no sense, since if
we compare both its summands on the intersection of their natural
domains we obtain zero. But it is obvious how to get rid notationally
of this ugly contortion, which makes violence to DiffRen.

Let us start in earnest by considering, as the authors in that paper
do, the example of the winecup graph or ice-cream ladder graph, devoid
of internal vertices:
$$
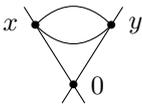
\begin{tikzpicture}
\coordinate (A) at (0,0) ; \coordinate (B) at (1,0) ; 
\coordinate (C) at (0.5,-0.8) ;
\draw ($ (A)!-0.3!(C) $) -- ($ (A)!1.3!(C) $) ;
\draw ($ (B)!-0.3!(C) $) -- ($ (B)!1.3!(C) $) ;
\draw (A) parabola[bend pos=0.5] bend +(0,0.25) (B) ;
\draw (A) parabola[bend pos=0.5] bend +(0,-0.25) (B) ;
\draw (A) node[left=3pt] {$x$} ;
\draw (B) node[right=3pt] {$y$} ;
\draw (C) node[right=3pt] {$0$} ;
\foreach \pt in {A,B,C} \draw (\pt) node{$\bull$} ;
\end{tikzpicture}
$$
(In order to get closer to the notation in
\cite{LCristinaMVV94}, we have exchanged vertices~$x$ and~$0$ in
the formulae in~\cite{Elara}.) We denote it $\copadevino(x,y)$ for
future use. The corresponding bare amplitude is given by
$$
f(x,y) = \frac{1}{x^2y^2(x - y)^4} \,.
$$

Both papers~\cite{FreedmanJL92} and~\cite{Elara} consider a
``partially renormalized'' version of the winecup graph, for which the
known formulas respectively yield:
\begin{align}
r_{x,y} \bigl[ x^{-2}y^{-2}(x - y)^{-4}\bigr] &= - \frac{1}{2}\,
x^{-2}y^{-2} \,\Dl\Bigl((x - y)^{-2} \log \frac{|x - y|}{l} \, \Bigr);
\label{eq:prematur} 
\\
R_{x,y} \bigl[ x^{-2}y^{-2}(x - y)^{-4}\bigr] &= - \frac{1}{2}\,
x^{-2}y^{-2} \,\Dl\Bigl((x - y)^{-2} \log \frac{|x - y|}{l} \, \Bigr)
\nonumber \\
&\hspace*{1.2em}  + \pi^2 x^{-4}\,\dl(x - y).
\notag 
\end{align}
The last expressions indeed make sense for all $(x,y)\neq(0,0)$. 

The first term on the right hand sides above is the one given in
\cite[Eq.~2.15]{LCristinaMVV94}. Extension of the last term, not
present there, to the thin diagonal (i.e., the whole graph), clearly is
no problem. So we concentrate in extending the first: each of the
factors in $r_{x,y}\bigl[x^{-2}y^{-2}(x - y)^{-4}\bigr]$ is a
well-defined distribution, but their product is not.

The tactic followed in~\cite{FreedmanJL92,Elara} is to invoke Green's
integration-by-parts formula to shift the Laplacian to the left and
use the fundamental solution for it. Paper \cite{LCristinaMVV94}
purports instead to deal with the overall divergence ``as a
three-point problem''. Their claim is firstly that the whole graph is
renormalized by
$$
(I - T_{x,y,0})(I - T_{x,0})x^{-2}y^{-2}(x - y)^{-4}.
$$
Secondly, that the handy and correct formula
\begin{equation}
A(\Dl B) = (\Dl A)B + \del_\bt(A \del_\bt B - B \del_\bt A),
\label{eq:aku-byodo} 
\end{equation}
employed in~\cite{FreedmanJL92} and borrowed by~\cite{Elara}, is not to
be used, on the grounds that such a trick reverts to a two-point
problem, while they want to grapple directly with the three-point
problem. Thirdly, that the singular behaviors of
$r_{x,y}\bigl[ x^{-2} y^{-2} (x-y)^{-4} \bigr]$ and of
$-2\pi^2\,\dl(y)\,x^{-4}\log|x|/l$ are the same.

Again, the mathematical argument there given for all that is hard to
bear. However, the last assertion is correct, and it can be made sense
of as follows. Note that
\begin{align*}
\MoveEqLeft 
\Bigl< x^{-2} z^{-2}\,
\Dl\Bigl((x - z)^{-2} \log \frac{|x - z|}{l} \Bigr), \vf(x,z) \Bigr>
\\
&= \Bigl< x^{-2} z^{-2}\,
\Dl\Bigl((x - z)^{-2} \log \frac{|x - z|}{l} \Bigr), \vf(x,0) \Bigr>
\\
&\quad + \Bigl< x^{-2} y^{-2}
\,\Dl\Bigl((x - y)^{-2} \log\frac{|x - y|}{l} \Bigr),
\vf(x,y) - \vf(x,0) \Bigr> \,.
\end{align*}
The second integral is finite, and the first one is proportional to
$$
\duo< \dl(y)\, x^{-4} \log|x|/l, \vf(x,y)>.
$$
In practice we are back to the two-point problem. A similar argument
works whenever the graph has been rendered ``primitive'' by partial
renormalization. The authors of~\cite{LCristinaMVV94} continue
their exposition as follows. In DiffRen one has
$$
r_{x,0}\bigl[ x^{-4} \log|x|/l \bigr]
= -\frac{1}{4}\, \Dl\frac{\log^2|x|/l + \log|x|/l}{x^2}\,.
$$
Therefore they write:
\begin{align*}
& r_{x,y,0}\bigl[ r_{x,0}[x^{-2} y^{-2} (x - y)^{-4}] \bigr]
= - \frac{1}{2}\, x^{-2} y^{-2}\,
\Dl\bigl( (x - y)^{-2}\log|x - y|/l \bigr)
\\
&\qquad - 2\pi^2\,\dl(y) x^{-4}\log|x|/l
- \frac{\pi^2}{2}\,\dl(y) \,\Dl\frac{\log^2|x|/l + \log|x|/l}{x^2}\,,
\end{align*}
with the contention that, although neither of the first two terms in
the last expression exists \textit{strictu sensu}, their combination
is a well-defined distribution. The situation is then clarified by use
of~\eqref{eq:aku-byodo} nevertheless, whereby the second term in the
above display completely drops out, and we are left with the third one
plus a well-defined divergence ---in the sense of vector calculus, see
right below.

With our method, there comes
\begin{align*}
\MoveEqLeft 
- \frac{1}{2}\, x^{-2} y^{-2}\,
\Dl_y \Bigl( (x - y)^{-2} \log\frac{|x - y|}{l} \Bigr)
\\
&= 2\pi^2\,\dl(y) x^{-4} \log\frac{|x|}{l}
+ \frac{1}{2}\, x^{-2} \,\del_y^\bt L_\bt(x - y;y),
\end{align*}
where
$$
L_\bt(x - y;y) 
:= (x - y)^{-2}\log\frac{|x - y|}{l}\, \del^y_\bt y^{-2} 
- y^{-2}\,\del^y_\bt \Bigl( (x - y)^{-2} \log\frac{|x - y|}{l} \Bigr).
$$
Moreover, the term $2\pi^2\,\dl(y) x^{-4} \log|x|/l$ in our
treatment, as done already in~\eqref{eq:prematur}, is renormalized
according to \cite[Eq.~A.4]{Elara}:
$$
2\pi^2R_{x,y,0}[x^{-4}\log|x|/l\,\dl(y)] 
= -\frac{\pi^2}{2}\,\dl(y) \,\Dl \frac{\log^2|x|/l + \log|x|/l}{x^2}
+ \pi^4\,\dl(x)\,\dl(y).
$$
Finally, we ought to contend with the last term
in~\eqref{eq:prematur}:
$$
R_{x,y,0}[\pi^2 x^{-4}\,\dl(x - y)]
= -\frac{\pi^2}{2}\,\Dl \frac{\log|x|/l}{x^2} \,\dl(x - y)
+ \pi^4\,\dl(x)\,\dl(y).
$$

In summary, with obvious abbreviated notations:
\begin{align*}
R_{x,y,0} \copadevino(x,y) 
&= \frac{1}{2} x^{-2} \,\del_y^\bt L_\bt(x - y;y)
- \frac{\pi^2}{2}\, \dl(y)\, \Dl\frac{\log^2|x|/l + \log|x|/l}{x^2}
\\
&\qquad - \frac{\pi^2}{2}\, \Dl\frac{\log|x|/l}{x^2} \,\dl(x - y)
+ 2\pi^4\,\dl(x)\,\dl(y);
\\
r_{x,y,0} \copadevino(x,y)
&= \frac{1}{2} x^{-2} \,\del_y^\bt L_\bt(x - y;y)
- \frac{\pi^2}{2}\,\dl(y)\, \Dl\frac{\log^2|x|/l + \log|x|/l}{x^2}\,.
\end{align*}
We remark that the differences between DiffRen and the improved
Epstein--Glaser method are of no consequence for the $\beta$-function,
up to third order in the coupling constant~\cite[Sect.~6]{Elara}.

Note that we may as well symmetrize:
\begin{align*}
r_{x,y,0} \copadevino(x,y)
&= \frac{1}{4} x^{-2} \,\del_y^\bt L_\bt(x - y;y)
- \frac{\pi^2}{4}\,\dl(y)\, \Dl\frac{\log^2|x|/l + \log|x|/l}{x^2}
\\
&\quad + \frac{1}{4} y^{-2} \,\del_y^\bt L_\bt(x - y;x)
- \frac{\pi^2}{4}\,\dl(x)\, \Dl\frac{\log^2|y|/l + \log|y|/l}{y^2}\,.
\end{align*}

\section{More general procedures}
\label{sec:al-military}

It is plausibly claimed in \cite{LCristinaMVV94} that more
complicated graphs can be tackled by the systematics of the Bogoliubov
recursion, or its descendant the forest formula. That is,
\begin{equation}
R\Ga = (I - T_\Ga) \Rbar_\Ga,
\label{eq:Reaganomics1} 
\end{equation}
where the $\Rbar_\Ga$ object, renormalized but for the overall
divergence is given by:
\begin{equation}
\Rbar_\Ga 
= I + \sum_\sP \prod_{v\in\sP} (-T_v\Rbar v) \prod \mathrm{prop},
\label{eq:Reaganomics2} 
\end{equation}
where the sum is over the partitions of $\Ga$ into divergent
generalized vertices~$v$, and $\prod \mathrm{prop}$ denotes the
product of propagators corresponding to all lines which connect the
different elements of the partition.

The notational contortions above are again clear, but we do not take
issue with them; rather we hasten to revisit a definitely amusing
example. We ponder the six-loop graph $\Ga$ obtained when one ``stye''
develops on each of the single propagators in a winecup graph:
$$
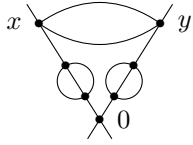
\begin{tikzpicture}[scale=0.8]
\coordinate (A) at (0,0) ; \coordinate (B) at (2,0) ; 
\coordinate (C) at (1.0,-1.6) ;
\coordinate (Ac) at ($ (A)!0.6!(C) $) ;
\coordinate (Bc) at ($ (B)!0.6!(C) $) ;
\coordinate (D) at ($ (A)!0.44!(C) $) ;
\coordinate (E) at ($ (A)!0.76!(C) $) ;
\coordinate (F) at ($ (B)!0.44!(C) $) ;
\coordinate (G) at ($ (B)!0.76!(C) $) ;
\draw ($ (A)!-0.2!(C) $) -- ($ (A)!1.2!(C) $) ;
\draw ($ (B)!-0.2!(C) $) -- ($ (B)!1.2!(C) $) ;
\draw (A) parabola[bend pos=0.5] bend +(0,0.35) (B) ;
\draw (A) parabola[bend pos=0.5] bend +(0,-0.35) (B) ;
\draw (Ac) circle(3mm) ; \draw (Bc) circle(3mm) ; 
\foreach \pt in {A,B,C,D,E,F,G} \draw (\pt) node{$\bull$} ;
\draw (A) node[left=3pt] {$x$} ;
\draw (B) node[right=3pt] {$y$} ;
\draw (C) node[right=3pt] {$0$} ;
\end{tikzpicture}
$$
Our authors claim that, after dealing with the subdivergences, the
partially renormalized graph comes out as
\begin{equation}
r_{x,y}\Ga \propto \frac{\log|x|/l}{x^2}\, \frac{\log|y|/l}{y^2}\,
\Dl\frac{\log|x - y|/l}{(x - y)^2} \,;
\label{eq:Abenomics} 
\end{equation}
and they set out to cure the overall divergence of the graph as
$x,y \to 0$. Our method sustains their claim: each of the ``dressed
propagator'' amplitudes for this diagram are of the form, with obvious
labels:
$$
\iint r_{x,0} \bigl[ v^{-2} (v - w)^{-6} (w - x)^{-2} \bigr] \,dv\,dw
= \iint r_{x,0} \bigl[ v^{-2} u^{-6} (v - u - x)^{-2} \bigr] \,du\,dv.
$$
Notice that this is a nested convolution; the inner integral is of the
form $R[r^{-6}] * r^{-2}$, which exists by the theory of
\cite[Sect.~3]{Elara}. As long as convolution can be effected, 
rule~\eqref{eq:in-annum} proceeds without obstruction.

The displayed integral becomes
$$
-\frac{1}{16} \iint v^{-2} (v - u - x)^{-2} \biggl( \Dl^2 \Bigl(
u^{-2} \log\frac{|u|}{l} \Bigr) - 5\pi^2 \,\Dl\dl(u) \biggr) \,dv\,du.
$$
On integrating by parts and dropping total derivatives in the
integrals over internal vertices, we then obtain
\begin{align*}
& \frac{\pi^2}{4} \iint v^{-2} \biggl( \Dl \Bigl( u^{-2}
\log\frac{|u|}{l} \Bigr) - 5\pi^2 \,\dl(u) \biggr) \,\dl(v - u - x)
\,dv\,du 
\\
&\quad = \frac{\pi^2}{4} \int (u + x)^{-2} \biggl( \Dl\Bigl( u^{-2}
\log\frac{|u|}{l} \Bigr) - 5\pi^2 \,\dl(u) \biggr) \,du
\\
&\quad = - \pi^4 \int u^{-2} \log\frac{|u|}{l} \,\dl(u + x)\,du
- \frac{5\pi^4}{4}\, x^{-2} 
= - \pi^4 x^{-2} \Bigl( \log\frac{|x|}{l} + \frac{5}{4} \Bigr),
\end{align*}
where the extra term with respect to formula~\eqref{eq:Abenomics} is
due to our different treatment of the basic ``sunset'' self-energy
diagram: as usual we define it so that
$$
x^2R_{x,0}[x^{-6}] = R_{x,0}[x^{-4}],
$$
which fails for DiffRen.

Therefore we have to renormalize the overall divergence
\begin{align*}
& \pi^8 \biggl( \frac{\log|x|/l}{x^2}\,\frac{\log|y|/l}{y^2}
+ \frac{5}{4} x^{-2} \frac{\log|y|/l}{y^2}
+ \frac{5}{4} y^{-2} \frac{\log|x|/l}{x^2}
+ \frac{25}{16} x^{-2} y^{-2} \biggr)
\\
&\qquad \x \biggl(- \frac{1}{2}\,
\Dl\Bigl( (x - y)^{-2} \log \frac{|x - y|}{l} \Bigr)
+ \pi^2\,\dl(x - y) \biggr).
\end{align*}
Several of the terms above bring nothing new; we concentrate on the
most difficult one, of the form~\eqref{eq:Abenomics}, the only one
recognized in \cite{LCristinaMVV94}. Its authors argue that as
$x \sim y \sim 0$ one has:
\begin{equation}
r_{x,y} \Ga \sim -32\pi^2 \,\dl(y)\, \frac{\log^3|x|/l}{x^4} \,,
\label{eq:kimono1} 
\end{equation}
which they renormalize en DiffRen as
\begin{equation}
4\pi^2\,\dl(y)\, \Dl\frac{\log^4|x|/l + 2\log^3|x|/l + 3\log^2|x|/l
+ 3\log|x|/l}{x^2}\,.
\label{eq:kimono2} 
\end{equation}
We pause to point to our essentially coincident formula
\cite[A.2]{Elara}:
\begin{align*}
\MoveEqLeft 
R_{x,0}\,\frac{\log^3|x|/l}{x^4} 
\\
&= - \frac{1}{8}\, \Dl \frac{\log^4|x|/l + 2\log^3|x|/l
+ 3\log^2|x|/l + 3\log|x|/l}{x^2} + \frac{3}{4}\,\pi^2\,\dl(x).
\end{align*}
Of course, the expression resulting from~\eqref{eq:kimono1} 
and~\eqref{eq:kimono2}:
\begin{align*}
& \frac{\log|x|/l}{x^2}\, \frac{\log|y|/l}{y^2}\,
\Dl\frac{\log|x - y|/l}{(x - y)^2}
+ 4\pi^2\,\dl(y) \biggl[ \frac{8\log^3|x|/l}{x^4}
\\
&\qquad + \Dl\frac{\log^4|x|/l + 2\log^3|x|/l + 3\log^2|x|/l
+ 3\log|x|/l}{x^2} \biggr]
\end{align*}
is rather ugly, since the first two terms are undefined. However, they
cleverly add and subtract to it $y^{-2}x^{-2}\log{|x|/l}\,\Dl\bigl((x
- y)^{-2}\log|x - y|/l\bigr)$, and applying Green's formula,
everything is rewritten:
\begin{align*}
\MoveEqLeft 
\frac{\log|x|/l}{x^2}\, \frac{\log|y|/l}{y^2}\,
\Dl\frac{\log|x - y|/l}{(x - y)^2}
\longmapsto \frac{\log|x|/l}{x^2}\, \frac{\log|y|/|x|}{y^2}\,
\Dl\frac{\log|x - y|/l}{(x - y)^2}
\\
& + \frac{\pi^2}{2}\,\dl(y) \biggl[ \Dl\frac{\log^4|x|/l
+ 2\log^3|x|/l + 3\log^2|x|/l + 3\log|x|/l}{x^2} \biggr]
\\
& - \frac{\log^2|x|/l}{x^2}\, \del_y^\bt L_\bt(x - y;y).
\end{align*}

The way to improve on this is symmetrization:
\begin{align*}
\MoveEqLeft 
r_{x,y,0} r_{x,y} \Ga
= \frac{\pi^2}{2}\,\dl(y) \biggl[ \Dl\frac{\log^4|x|/l
+ 2\log^3|x|/l + 3\log^2|x|/l + 3\log|x|/l}{x^2} \biggr]
\\
& + \frac{\pi^2}{2}\,\dl(x) \biggl[ \Dl\frac{\log^4|y|/l
+ 2\log^3|y|/l + 3\log^2|y|/l + 3\log|y|/l}{y^2} \biggr]
\\
& - \frac{\log^2|x|/l}{x^2}\, \del_y^\bt L_\bt(x - y;y)
- \frac{\log^2|y|/l}{y^2}\, \del_x^\bt L_\bt(y - x;x).
\end{align*}

The task of computing the remaining terms in $R_{x,y,0} R_{x,y}\,\Ga$
is comparatively easier. The ones containing the factor $\dl(x - y)$
simply go into
$$
\pi^{10} \dl(x - y) \biggl[ R_{x,0}\,\frac{\log^2|x|/l}{x^4}
+ \frac{5}{2}\, R_{x,0}\,\frac{\log|x|/l}{x^4}
+ \frac{25}{16}\, R_{x,0}\,\frac1{x^4} \biggr],
$$
in terms of known renormalized expressions~\cite{Elara}. Of the three
remaining terms, two are totally similar:
$$
- \frac{5}{8} \Bigl( x^{-2}\,\frac{\log|y|/l}{y^2}
+ y^{-2}\,\frac{\log|x|/l}{x^2} \Bigr)
\Dl\Bigl( (x - y)^{-2} \log\frac{|x - y|}{l} \Bigr),
$$
and can be computed as above by Green's formula, and the other is of
the same form as the winecup graph.

\section{Conclusion}
\label{sec:al-jaba}

Formulae \eqref{eq:Reaganomics1} and~\eqref{eq:Reaganomics2} work like
guiding principles, rather than calculational recipes. Actual
production of closed formulas relies on a bag of tricks. While the
application of the inductive principle of~\cite{Elara} often profits
from similar tricks, it appears to be better adapted in practice to
deal with complex diagrams.

\subsection*{Acknowledgments}
The author is thankful to Joseph C. V\'arilly for some very useful 
comments. His work was supported by the Spanish Ministry for Science
through grant FPA2012--35453.

\end{document}